\documentclass[lettersize,journal]{IEEEtran}
\usepackage{amsmath,amsfonts,amssymb}
\usepackage{algorithmic}
\usepackage{algorithm}
\usepackage{array}
\usepackage[caption=false,font=scriptsize,labelfont=rm,textfont=rm]{subfig}
\usepackage{textcomp}
\usepackage{stfloats}
\usepackage{url}
\usepackage{verbatim}
\usepackage{graphicx}
\usepackage{cite}
\usepackage[numbers,sort&compress]{natbib}

\usepackage{lettrine}
\usepackage{hyperref}
\usepackage{makecell}
\usepackage{enumitem}
\usepackage{svg}
\usepackage{cleveref}
\crefname{figure}{fig.}{figures}
\Crefname{figure}{Fig.}{Figures}
\usepackage{indentfirst}
\usepackage{float}
\usepackage{subfloat}
\usepackage{makecell}

\begin{document}


\title{Collaborative Channel Access and Transmission for NR Sidelink and Wi-Fi Coexistence over Unlicensed Spectrum}

\author{Zhuangzhuang~Yan,~Xinyu Gu,~\IEEEmembership{Member, IEEE},~Zhenyu Liu,~\IEEEmembership{Member, IEEE}, Liyang Lu
\thanks{This work was supported in part by the National
Key Research and Development Program under Grant 2022YFF0610303, in part by the Beijing Natural Science Foundation under Grant L242088, and in part by the National Natural Science Foundation of China (NSFC) under Grant 62201089.}
\thanks{Zhuangzhuang Yan and Xinyu Gu are with the School of Artificial Intelligence, Beijing University of Posts and Telecommunications, Beijing 100876, China (email: yanzz@bupt.edu.cn,  guxinyu@bupt.edu.cn)}
\thanks{Xinyu Gu is with the Purple Mountain Laboratories, Nanjing 211111, China}
\thanks{Zhenyu Liu is with the 5GIC and 6GIC, Institute for Communication Systems, University of Surrey, United Kingdom, Guildford, U.K.(email: zhenyu.liu@surrey.ac.uk)
}
\thanks{Liyang Lu is with the State Key Lab of Intelligent Transportation System, Beijing 100088, China(email: lly@itsc.cn)
}
}


\sloppy
\maketitle

\begin{abstract}
With the rapid development of various internet of things (IoT) applications, including industrial IoT (IIoT) and visual IoT (VIoT), the demand for direct device-to-device communication to support high data rates continues to grow. To address this demand, 5G-Advanced has introduced sidelink communication over the unlicensed spectrum (SL-U) to increase data rates. However, the primary challenge of SL-U in the unlicensed spectrum is ensuring fair coexistence with other incumbent systems, such as Wi-Fi. In this paper, we address the challenge by designing channel access mechanisms and power control strategies to mitigate interference and ensure fair coexistence. First, we propose a novel collaborative channel access (CCHA) mechanism that integrates channel access with resource allocation through collaborative interactions between base stations (BS) and SL-U users. This mechanism ensures fair coexistence with incumbent systems while improving resource utilization. Second, to further enhance the performance of the coexistence system, we develop a cooperative subgoal-based hierarchical deep reinforcement learning (C-GHDRL) algorithm framework. The framework enables SL-U users to make globally optimal decisions by leveraging cooperative operations between the BS and SL-U users, effectively overcoming the limitations of traditional optimization methods in solving joint optimization problems with nonlinear constraints. Finally, we mathematically model the joint channel access and power control problem and balance the trade-off between fairness and transmission rate in the coexistence system by defining a suitable reward function in the C-GHDRL algorithm. Simulation results demonstrate that the proposed scheme significantly enhances the performance of the coexistence system while ensuring fair coexistence between SL-U and Wi-Fi users.
\end{abstract}

\begin{IEEEkeywords}
SL-U, collaborative, hierarchical deep reinforcement learning, channel access, power control.
\end{IEEEkeywords}

\section{Introduction}
\IEEEPARstart{W}{ith} the advent of industry 5.0, driven by emerging applications such as predictive maintenance, hyper-customization in manufacturing, and collaborative robotics \cite{ref1}, the data rates of future Industrial Internet of Things (IIoT) systems are predicted to increase by 10 to 50 times compared to current networks \cite{ref2}. Additionally, the Visual Internet of Things (VIoT) plays a crucial role in urban counter-terrorism, intelligent border security, and missing person searches by leveraging visual sensors to process and analyze real-time visual data streams intelligently \cite{ref3}. Either the predictive maintenance within IIoT or the transmission of video streams in VIoT, sidelink (SL) communication with high data rates is required to deliver these services effectively. To address the demand, the evolution of new radio (NR) sidelink \cite{ref4}, one of the initial projects approved by 5G-Advanced, offers a solution by utilizing sidelink over unlicensed spectrum (SL-U). SL-U combines sidelink and NR on unlicensed channel spectrum (NR-U) technologies \cite{addref44}\cite{ref7}, leveraging idle unlicensed spectrum to enhance the data rates. So, it is highly suitable for short-distance wireless communication applications like the Internet of Things (IoT) \cite{addref36}.

While the introduction of SL-U can significantly enhance the data rate of wireless links, the primary challenge for SL-U is to ensure fair coexistence with other incumbent systems, such as Wi-Fi, in the unlicensed spectrum. The most critical of fair coexistence is the need for effective interference mitigation to ensure fair and efficient spectrum sharing \cite{addref37}\cite{addref38}. To address the challenge, SL-U employs the listen-before-talk (LBT) channel access mechanism, which enables devices to perform a clear channel assessment (CCA) to sense channel availability prior to transmission \cite{ref5}. Although the LBT mechanism effectively mitigates interference by selecting idle channels, the existing LBT channel access mechanism may lead to channel collisions and low resource utilization for SL-U users due to slot alignment \cite{ref7}\cite{ref8}. To address this issue, \cite{ref7} proposed a method to reduce collisions by randomizing the starting position of the back-off countdown process within a slot. However, the approach does not account for coexistence scenarios. \cite{ref8} proposed a new CR-LBT (LBT With Collision Resolution) method of transmitting reserved signals, which greatly reduces collisions and enhances the fairness of channel resource sharing. However, the presence of reserved signals wastes channel resources.


Although the LBT mechanism effectively mitigates interference, efficient resource reuse in specific coexistence scenarios can further enhance spectrum utilization. During the process, power control is essential for effectively mitigating interference \cite{ref9}\cite{addref39}\cite{ref10}\cite{ref11}. Considering inter-cell and inter-operator interference in both licensed and unlicensed spectrum, \cite{ref10} proposed a joint optimization framework that integrates user association, power allocation, and dynamic spectrum sharing to maximize network throughput. \cite{ref11} presented a joint power control and time allocation scheme for licensed and unlicensed spectrum in unmanned aerial vehicle (UAV)-assisted Internet of Vehicles (IoV) systems, optimizing power and time allocation to maximize overall system capacity.

Therefore, channel access and power control are two fundamental mechanisms for SL-U in the coexistence network, and they exhibit significant interdependence \cite{ref7}\cite{ref9}\cite{ref12}. This interdependence becomes especially significant in unlicensed spectrum coexistence scenarios. For example, considering the coexistence of terrestrial cellular networks, Wi-Fi, and UAV-based base stations (UAV-BSs), \cite{ref13} jointly optimized sub-channel allocation and power control of UAV users across the licensed and unlicensed spectrum to maximize the total uplink rate. \cite{ref14} jointly optimized the time and power allocation during maximum channel occupation time (MCOT) to maximize the total throughput of both downlink and uplink in NR-U while ensuring fair coexistence with Wi-Fi. Various centralized algorithms based on combinatorial optimization \cite{ref15}, random geometry \cite{ref16}, and game theory \cite{ref17} have been proposed for the joint optimization problem. However, the joint optimization of channel access and power control is often formulated as a combinatorial optimization problem with nonlinear constraints, which leads to the ineffectiveness of traditional optimization methods in most scenarios \cite{ref9}\cite{ref18}.

The emergence of deep reinforcement learning (DRL) has introduced promising solutions to joint optimization problems with nonlinear constraints \cite{ref19}\cite{addref42}\cite{ref20}\cite{ref21}. In \cite{ref20}, a distributed DRL-based scheme was proposed, enabling D2D pairs to autonomously optimize channel selection and transmission power using only local information and outdated non-local information. \cite{ref21} introduced a novel DRL-based framework for jointly optimizing power and block length allocation to minimize the worst-case decoding-error probability in the finite block length (FBL) regime for an ultra-reliable low-latency communication (URLLC) -based downlink V2X communication system. However, it is important to note that DRL-based approaches typically require all actions to operate on the same time scale when solving joint optimization problems, meaning that all actions must be decided simultaneously. This assumption may not be applicable in certain communication scenarios where actions occur at different timescales.

Therefore, hierarchical deep reinforcement learning (HDRL) is proposed to solve the problem of joint optimization that typically requires all operations to be at the same time scale \cite{ref23}\cite{ref24}\cite{addref43}\cite{ref25}\cite{ref26}. By adopting a hierarchical learning framework, HDRL can achieve both global and local optimization, which is very suitable for the joint optimization of channel access and power control with different timescales. In \cite{ref25}, the authors optimized the CCA threshold and transmit power of the access point (AP) using the HDRL algorithm, improving network performance. However, \cite{ref25} focuses solely on a scenario involving Wi-Fi users and does not address the coexistence challenges associated with the unlicensed spectrum. \cite{ref26} proposed leveraging the HDRL algorithm to jointly optimize partial spectrum sensing and power allocation, enabling users to make intelligent decisions on spectrum selection and power allocation to maximize network throughput while minimizing mutual interference. Although each secondary user (SU) transmits information to the central server to train the HDRL neural network, the central server maintains a separate model for each user. As a result, each network is trained based solely on the local information available to the corresponding user, which may prevent the model from achieving a globally optimal solution.

In summary, we propose a novel collaborative channel access (CCHA) mechanism to address channel collisions and low resource utilization for SL-U users. Although the mechanism can improve resource utilization, effective resource reuse can further enhance resource utilization in some coexistence scenarios. Therefore, we jointly optimize channel access and power control to maximize spectrum efficiency. To overcome the inefficiencies of traditional combinatorial optimization methods under nonlinear constraints and joint optimization across different time scales, we developed a collaborative subgoal-based HDRL (C-GHDRL) algorithm framework. The framework achieves joint optimization of channel access and power control through collaborative interactions between the BS and SL-U users, thereby significantly improving the overall performance of the coexistence system. Finally, we model the joint channel access and power control problem and achieve a balance between fairness and transmission rate by defining an appropriate reward function within the algorithm. The main contributions of this paper are as follows:
\begin{itemize}
\item We propose a CCHA mechanism integrating channel access and resource allocation to ensure fair coexistence. In this mechanism, the BS allocates available resources to SL-U users within its coverage area based on channel sensing results. SL-U users assigned to these resources perform lightweight LBT before accessing the channel. This collaborative interaction between the BS and SL-U users not only ensures complete resource orthogonality but also increases the opportunities for channel access for SL-U users. Consequently, this mechanism enhances resource utilization and significantly improves coexistence system performance.

\item To further enhance the coexistence system performance, we propose a C-GHDRL algorithm framework designed to jointly optimize channel access and power control for SL-U users in coexistence systems. In the framework, BS collects global information through interaction with SL-U users to train the C-GHDRL model and periodically distributes the trained model to all SL-U users. Each SL-U user applies this model autonomously to determine optimal channel access and transmission power. Through collaborative operations between the BS and SL-U users, the algorithm enables SL-U users to make globally optimal decisions while effectively addressing the joint optimization of channel access and power control under nonlinear constraints, thereby significantly improving the overall performance of coexistence systems.

\item We formulate a mathematical joint model for the channel access and power control problems as the joint optimization goal. By defining an appropriate reward function within the C-GHDRL algorithm, we balance fairness and transmission rate in the coexistence system.

\end{itemize}

The rest of this paper is organized as follows. Section II presents the system model as well as the problem formulation. In Section III, we elaborate on the proposed collaborative channel access mechanism. Joint optimization of channel access and power control based on the C-GHDRL algorithm is described in Section IV. Followed by performance evaluation and comparison in Section V. Finally, a conclusion is drawn in Section VI.

\begin{figure}[t]
\centering
\includegraphics[width=0.45\textwidth]{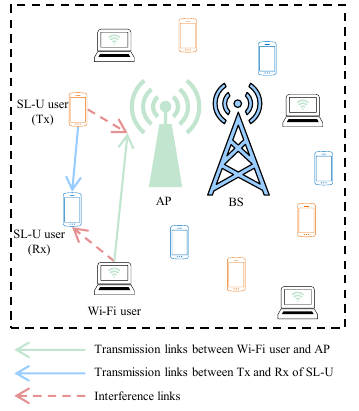}
\caption{Wi-Fi and SL-U coexistence scenario.}
\label{fig:NM}
\end{figure}

\section{System Model and Problem Formulation}
\begin{description}
\item[A.] Network Model
\end{description}

Consider a communication scenario in which Wi-Fi and SL-U coexist in the unlicensed spectrum, as depicted in \Cref{fig:NM}. This scenario includes a Wi-Fi AP, an SL-U BS, multiple Wi-Fi users, and multiple SL-U user pairs. Each SL-U user pair comprises a transmitter (Tx) and a receiver (Rx). The BS and AP are centrally located within the area, while Wi-Fi and SL-U users are randomly distributed. Let $ \mathcal{M} = \{1,2,\cdots,M\} $ denote the set of SL-U user pairs, and $ \mathcal{N} = \{1,2,\cdots,N\} $ denote the set of Wi-Fi users. Define $T_m$ and $R_m$ as the transmitter and receiver of SL-U user pair $m \in \mathcal{M}$ , respectively.

\begin{description}
\item[B.] Channel Access Schemes over Unlicensed Spectrum
\end{description}

Both SL-U and Wi-Fi users compete for access to the unlicensed spectrum using contention-based access mechanisms. In the IEEE 802.11n protocol, Wi-Fi users used Carrier Sense Multiple Access with Collision Avoidance (CSMA/CA) to manage channel contention \cite{ref27}. SL-U employed the LBT mechanism specified by 3GPP 37.213 \cite{ref5}. In the following, we will provide a detailed description of these two channel access mechanisms.

\begin{figure}[!t]
\centering
\includegraphics[width=0.48\textwidth]{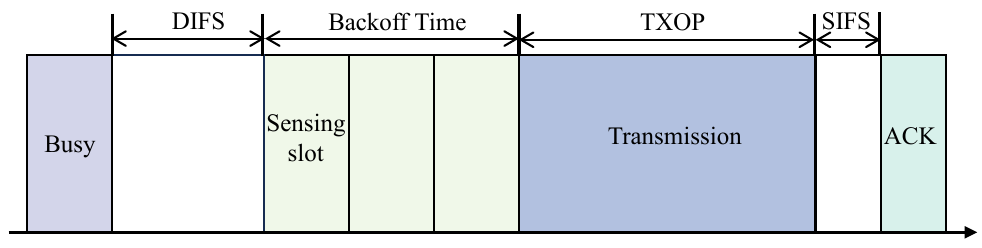}
\caption{CSMA/CA procedure of Wi-Fi.}
\label{fig:CSMA}
\end{figure}

\begin{enumerate}[label=\arabic*)] 
\item \textbf{CSMA/CA:} As illustrated in \Cref{fig:CSMA}, the CSMA/CA scheme incorporates a back-off mechanism to randomize the transmission start time of Wi-Fi users, thereby reducing the likelihood of collisions in the absence of a centralized coordination function. When a new data packet arrives, Wi-Fi users perform carrier sensing to check the status of the wireless channel before initiating transmission. When the user senses that the wireless channel has been idle for distributed inter-frame space (DIFS), it will proceed to wait for a random time before sending data. The period of random waiting is called the random back-off time, and the operation of waiting for a random time is called the random back-off process. Once the back-off period concludes, the Wi-Fi user transmits their data during a transmission opportunity (TXOP). Upon successfully receiving the data, the AP waits for a short inter-frame space (SIFS) and sends an acknowledgment (ACK) to confirm the correct reception.

\item \textbf{LBT:} The LBT scheme adopts a similar back-off mechanism to that of the CSMA/CA protocol. Before initiating data transmission, an SL-U user performs CCA to sense whether the channel is busy or idle. If the channel remains idle for a specified period, denoted as $T_d$, the user initiates a back-off process, introducing a random time before attempting transmission. \cite{ref5} defined two types of channel access processes for LBT:
\begin{itemize}
\item \textbf{Type 1:} Channel access procedure with random back-off of variable size contention window (CW). According to \cite{ref5}, the CW size is determined by the channel access priority $p$, as indicated in Table \ref{table:CAPC}. For example, when $p = 1$, the time a user requires to sense the channel is given by $T_{sensing} = T_d + m_p \times T_{sl} + CW_p \times T_{sl} = 34 \mu s + CW_p \times 9\mu s$, where $T_{sl}$ is sensing slot.
\item \textbf{Type 2:} No random backoff procedure is performed, and channel sensing is conducted for a duration of $T_{short\_sl} = 16\ or\ 25 \mu s$ before accessing the channel. Notably, the channel sensing duration for Type 2 is shorter than that of Type 1. 
\end{itemize}
\end{enumerate}

\begin{table*}[!t]
\center
\caption{Channel Access Priority Class (CAPC)} 
\scriptsize
\renewcommand{\arraystretch}{1}
\resizebox{0.9\textwidth}{!}{
\begin{tabular}{|c|c|c|c|c|c|} 
\hline
\makebox[0.15\textwidth][c]{\makecell[c]{Channel Access \\ Priority Class $(p)$}} & $m_p$ & $CW_{min,p}$ & $CW_{max,p}$ & $T_{mcot,p}$ &  allowed $CW_p$ sizes \\
\hline
1 & 2 & 3 & 7 & 2 ms & \{3,7\} \\ 
\hline
2 & 2 & 7 & 15 & 4 ms & \{7,15\} \\ 
\hline
3 & 3 & 15 & 1023 & 6 or 10 ms & \{15,31,63,127,255,511,1023\} \\ 
\hline
4 & 7 & 15 & 1023 & 6 or 10 ms & \{15,31,63,127,255,511,1023\} \\ 
\hline
\end{tabular}
}
\label{table:CAPC}
\end{table*}

\begin{description}
\item[C.] Communication Model
\end{description}

The Wi-Fi users and SL-U users path loss model is based on the 3GPP standard and applies to line-of-sight (LOS) scenarios\cite{ref28}. The path loss model is as follows:
\begin{equation}
PL_{w} = 32.4 + 20 \times \log_{10} {f_c} + 17.3 \times \log_{10} {(d_{w})}, \label{wifi_path_loss}
\end{equation}
\begin{equation}
PL_{s} = 32.4 + 20 \times \log_{10} {f_c} + 17.3 \times \log_{10} {(d_{s})}, \label{slu_path_loss}
\end{equation}
where $f_c$ is the center frequency in GHz.The $d_{w}$ is the distance between the Wi-Fi user and the AP in meters, and the $d_{s}$ is the distance between the transmitter and receiver of the SL-U user in meters.

Therefore, the signal-to-interference-to-noise ratio (SINR) of the SL-U pair $m$ and the Wi-Fi user $n$ in slot $t$ are respectively defined as:
\begin{equation}
\eta_{s,m}(t) = \frac {p_{s,m}^{T}(t) - PL_{s,m}}{p_{noise} + \sum\limits_{n=0}^{N}{k_{w,n}(t)p_{inter,w}^{R_m,n}(t)}}\label{slu_sinr},
\end{equation}
\begin{equation}
\eta_{w,n}(t) = \frac {p_{w,n}^{T}(t) - PL_{w,n}}{p_{noise} + \sum\limits_{m=0}^{M}{k_{s,m}(t)p_{inter,s}^{AP,m}(t)}}\label{wifi_sinr},
\end{equation}
where $p_{s,m}^{T}(t)$ and $p_{w,n}^{T}(t)$ are the transmission power of SL-U user $T_m$ and the Wi-Fi user $n$, respectively. The $PL_{s,m}$ represents the path loss between the transmitter and receiver of the SL-U pair $m$. The $PL_{w,n}$ is the path loss of the Wi-Fi user $n$. The $p_{noise}$ is additive Gaussian white noise (AWGN) power, $p_{inter,w}^{R_m,n}(t)$ represents the interference power received by SL-U $R_m$ from Wi-Fi user $n$, $p_{inter,s}^{AP,m}(t)$ represents the interference power received by AP from SL-U $T_m$. If $k_{s,m}(t) = 1$ indicates that SL-U pair $m$ accesses the channel during slot $t$; otherwise, $k_{s,m}(t) = 0$ indicates that SL-U pair $m$ does not access the channel. If $k_{w,n}(t) = 1$ indicates that Wi-Fi user $n$ accesses the channel during slot $t$; otherwise, $k_{w,n}(t) = 0$ indicates that Wi-Fi user $n$ does not access the channel.

According to \eqref{slu_sinr}\eqref{wifi_sinr}, the transmission rate of the SL-U system and Wi-Fi system in slot $t$ are respectively defined as:
\begin{equation}
R_s^C(t) = \sum\limits_{m=0}^{M}{k_{s,m}(t)B\log_{2}{(1 + \eta_{s,m}(t))}}\label{slu_Rc},
\end{equation}
\begin{equation}
R_w^C(t) = \sum\limits_{n=0}^{N}{k_{w,n}(t)B\log_{2}{(1 + \eta_{w,n}(t))}}\label{wifi_Rc},
\end{equation}
where the $B$ denotes channel bandwidth.

Therefore, the total transmission rate of the coexistence system in slot $t$ is defined as:
\begin{equation}
R_{total}^C(t) = R_s^C(t) + R_w^C(t)\label{total_Rc},
\end{equation}

\begin{description}
\item[D.] Fairness Model Analysis
\end{description}

Fairness measure is used in network engineering to determine whether users or applications are receiving a fair share of system resources. The formula for Jain’s fairness index is as follows \cite{ref29}:
\begin{equation}
\mathcal{J}(x_1,x_2,\cdots,x_n) = \frac{(\sum\limits_{i=1}^{n}{x_i})^2}{n\cdot \sum\limits_{i=1}^{n}{x_i^2}} \in [\frac{1}{n}, 1]\label{Jain fairness index},
\end{equation}
where $x_i$ is the throughput for the $i$th connection. $n$ denotes the total number of users. According to the characteristics of Jain’s fairness index, it is a continuous value from $1/n$ to $1$. The index is $1$ when all users receive the same allocation. 

Although the SL-U system is authorized to transmit data in the unlicensed spectrum, it should not affect the performance of the Wi-Fi system. In this paper, we employ the Jain's fairness index, as described in \eqref{Jain fairness index}, to ensure fairness in the coexistence system, which is expressed as follows:
\begin{equation}
\mathcal{JF}^{C} = \frac{(R_s^C + R_w^C)^2}{2((R_s^C)^2 + (R_w^C)^2)},
\label{JF of Th}
\end{equation}
where $R_s^C$ and $R_w^C$ are the transmission rate attained by SL-U and Wi-Fi system during the simulation time, respectively. 

\begin{description}
\item[E.] Problem Formulation
\end{description}

In unlicensed spectrum coexistence systems, our objective is to maximize the transmission rate while maintaining fairness across the system. To comprehensively evaluate both the transmission rate and fairness between SL-U and Wi-Fi, we define the utility function as follows:
\begin{equation}
U(k,p) = \beta R_{total}^C + (1 - \beta)\mathcal{JF}^C,
\label{UF}
\end{equation}
where $\beta$ is a weighted factor to balance the transmission rate and the fairness.

We find that the maximizing utility function is closely related to channel access status $k$ and transmission power $p$. Therefore, the mathematical formulation of the problem is given as follows:
\begin{equation}
    \begin{aligned}
    &P.F.: \mathop{\arg\max}_{\boldsymbol{k},\boldsymbol{p}} U(\boldsymbol{k},\boldsymbol{p}),  \label{max of UF}\\
    & \begin{array}{r@{\quad}l@{}l@{\quad}l}
    & \\
    s.t.& C1 : p_{s,min}^T \leq p_{s,m}^T(t) \leq p_{s,max}^T, \forall m \in \mathcal{M},\\
        & \\
        & C2 : p_{w,min}^T \leq p_{w,n}^T(t) \leq p_{w,max}^T, \forall n \in \mathcal{N},\\
        & \\
        & C3 : \sum\limits_{m=0}^{M}{k_{s,m}(t)} \leq 1, \forall m \in \mathcal{M},\\
        & \\
        & C4 : \sum\limits_{n=0}^{N}{k_{w,n}(t)} \leq 1, \forall n \in \mathcal{N}.\\
        & \\
        & C5 : \sum\limits_{m=0}^{M}{k_{s,m}(t)} + \sum\limits_{n=0}^{N}{k_{w,n}(t)} \leq 2, \forall m \in \mathcal{M}, n \in \mathcal{N}.\\
    \end{array}  
\end{aligned}
\end{equation}
where constraints $C1$ and $C2$ restrict the transmission power of each SL-U user pair and Wi-Fi user, respectively. Constraints $C3$ indicates that only one SL-U user pair occupies the channel per slot at most. Constraint $C4$ shows that at most one Wi-Fi user occupies the channel per slot. Constraint $C5$ indicates that at most one Wi-Fi user and one SL-U user can use the channel simultaneously per slot.

\begin{figure}[t]
    \centering
    \subfloat[Collisions on the same channel.]{
        \centering
        \includegraphics[width=0.45\textwidth]{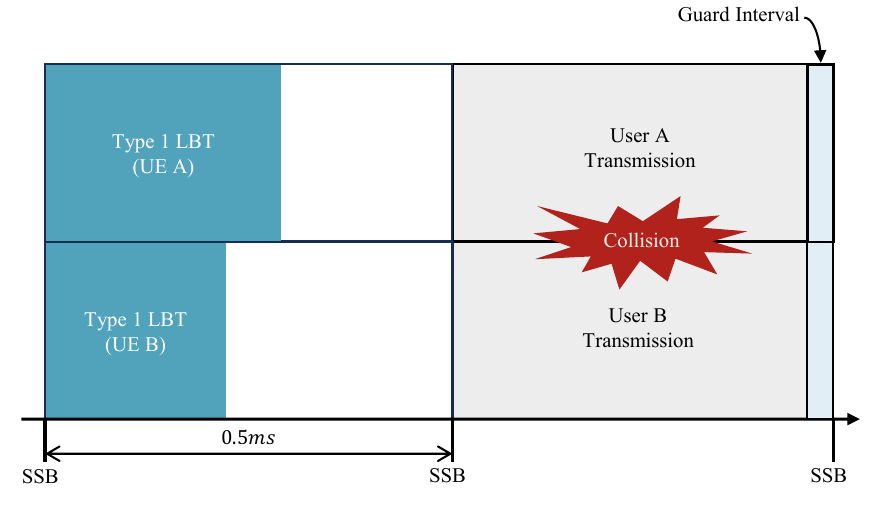} 
        \label{fig:CCHA_1}
    }
    \hfill 
    \subfloat[Channel occupancy based on centralized resource allocation and Tpye 1 LBT.]{
        \centering
        \includegraphics[width=0.45\textwidth]{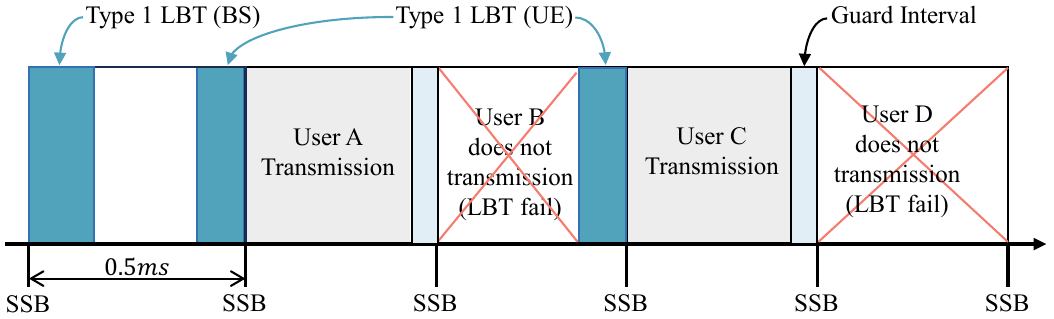} 
        \label{fig:CCHA_2}
    }
    \hfill 
    \subfloat[Channel occupancy based on collaborative channel access mechanism.]{
        \centering
        \includegraphics[width=0.45\textwidth]{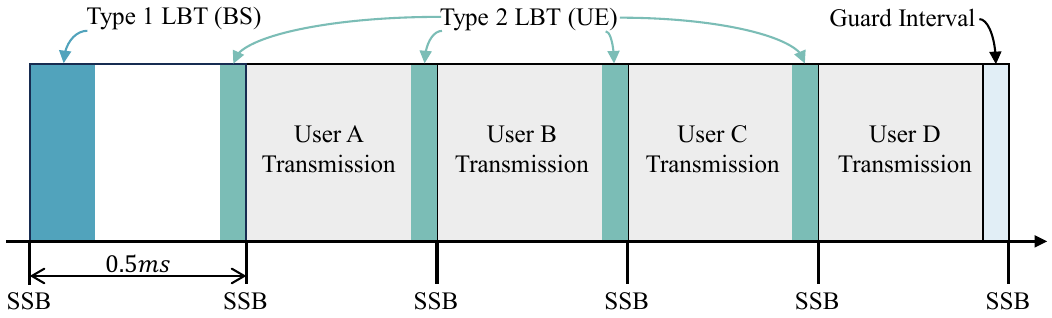} 
        \label{fig:CCHA_3}
    }
    \caption{Channel occupancy due to different schemes.}
    \label{fig:Channel occupancy}
\end{figure}

\section{Collaborative Channel Access (CCHA)}
In this section, we present a comprehensive explanation of the collaborative channel access mechanism. Furthermore, we describe the information acquisition process that supports these collaborative operations. 

SL-U users operating in the unlicensed spectrum cannot immediately begin transmission after completing the LBT process; instead, they must wait until the next spectrum slot boundary (SSB) to transmit data. This limitation significantly impacts the performance of SL-U systems. In a distributed resource allocation scheme, namely Mode 2, each SL-U user independently performs the LBT process and waits for the next SSB to transmit. However, this approach often results in frequent and severe collisions, as illustrated in \Cref{fig:Channel occupancy}\subref{fig:CCHA_1}. In the figure, SL-U users A and B independently perform Type 1 LBT channel sensing at specified locations. Although they sense the channel for different durations, they both wait for the next SSB to transmit data. And, since users A and B cannot know each other's channel occupancy status in the distributed system, they transmit data in the same slot, resulting in collision.

On the other hand, if SL-U users adopt a centralized resource allocation scheme, the collision issue can be mitigated. Nevertheless, if only Type 1 LBT is employed, the LBT process may fail due to the channel occupied in the previous slot by other SL-U users, resulting in low channel resource utilization, as shown in \Cref{fig:Channel occupancy}\subref{fig:CCHA_2}. In \Cref{fig:Channel occupancy}\subref{fig:CCHA_2}, SL-U users A, B, C, and D are assigned defined slots to transmit data. To comply with unlicensed spectrum occupancy rules, users must sense the channel at specified locations. Since user A occupies the channel, user B senses the channel as busy during channel sensing, preventing user B from transmitting data in the defined slot. The same applies to user D. This results in low resource utilization.

To address these issues, we propose a collaborative channel access mechanism designed to optimize resource utilization and enhance the overall performance of coexistence systems. This mechanism integrates multiple LBT types with a centralized resource allocation strategy to improve efficiency and coordination. First, the BS performs the Type 1 LBT process to compete for channel access. Upon successful completion, the BS obtains the channel's time-domain occupation period, denoted as $T_{cot}$, and the channel bandwidth $B$ in the frequency domain. The BS then centrally allocates these resources, dividing them into slots for SL-U users with data to transmit within its coverage area. SL-U users allocated specific channel resources confirm their availability by performing a Type 2 LBT process at designated locations before the SSB. According to the standard, SL-U users avoid transmitting data during the final orthogonal frequency division multiplexing (OFDM) symbol of the allocated resources to establish a guard interval (GI). For a subcarrier spacing of 30 kHz, this guard interval is 33.33 $\mu s$, which is sufficient to complete the Type 2 LBT process, with a sensing time of at most 25 $\mu s$. If the Type 2 LBT process is completed, the user proceeds with data transmission, as illustrated in \Cref{fig:Channel occupancy}\subref{fig:CCHA_3}. The mechanism avoids collisions, as shown in \Cref{fig:Channel occupancy}\subref{fig:CCHA_1}, by enabling the BS to allocate resources centrally. Furthermore, by using only the Type 2 LBT process, the channel sensing time is significantly shorter than the Type 1 LBT process, helping to prevent resource wastage, as seen in \Cref{fig:Channel occupancy}\subref{fig:CCHA_2}. By improving resource utilization, the mechanism enhances the performance of the coexistence system.

When implementing the collaborative channel access mechanism proposed in this study, the base station and SL-U users exchange control signals to facilitate collaborative operations. To ensure the reliability of control signaling transmission, these signals are conveyed over the licensed spectrum.



\section{Joint Optimization of Channel Access and Power Control based on C-GHDRL}
In this section, we first present a brief background of semi-Markov decision processes and hierarchical reinforcement learning. Subsequently, we introduce the C-GHDRL algorithm, designed to achieve the joint optimization of channel access and power control.

\begin{description}
\item[A.] SMDP and Hierarchical Reinforcement Learning
\end{description}

The Markov Decision Process (MDP) is commonly employed to model scenarios in reinforcement learning (RL), primarily addressing traditional decision-making processes. However, in real-world scenarios, the intervals between decision-making stages can be variable, which poses challenges for standard MDPs in accurately modeling such problems. The Semi-Markov Decision Process (SMDP), as an extension of the MDP framework, offers an option-based decision-making model capable of addressing control problems characterized by variable decision times \cite{ref30}.

The Semi-Markov Decision Process is primarily defined by a five-tuple $<\mathcal{S}, \mathcal{A}, \mathcal{O}, \mathcal{P}, \mathcal{R}>$, where $\mathcal{O}$ denotes the set of options, and the remaining elements have meanings analogous to those in the standard MDP \cite{ref31}. An option is represented as a triplet $<\mathcal{I}, \pi, \beta>$, where $\mathcal{I} \subseteq \mathcal{S}$ indicates the set of initial states for the option, $\pi:\mathcal{S} \times \mathcal{A} \rightarrow [0,1]$ represents the policy function associated with the option, and $\beta: \mathcal{S} \rightarrow [0,1]$ denotes the probability of the option terminating in a given state. An option $<\mathcal{I}, \pi, \beta>$ is available in state $s_t$ if and only if $s_t \in \mathcal{I}$. Once the option is selected, actions are determined according to $\pi$ until the option terminates stochastically based on $\beta$. Specifically, the execution of the Markov option proceeds as follows. First, the agent selects the next action $a_t$ according to the probability distribution $\pi(s_t, \cdot)$. The environment then transitions to state $s_{t+1}$, where the option may terminate with probability $\beta(s_{t+1})$. If the option does not terminate, the agent determines the action $a_{t+1}$ based on $\pi(s_{t+1}, \cdot)$ and possibly terminates in $s_{t+2}$ according to $\beta(s_{t+2})$, and so on. Upon termination of the option, the agent can choose another option.

Hierarchical Reinforcement Learning (HRL) is a significant subfield of reinforcement learning that distinguishes itself from classic RL by being grounded in the SMDP framework. HRL enhances the traditional RL approach through hierarchical abstraction, addressing challenges like sparse rewards, sequential decision-making, and limited transferability, which are often difficult to resolve in conventional RL. By structuring decision processes hierarchically, HRL promotes more effective exploration and improved transfer learning capabilities. Despite these advantages, HRL still faces limitations in computational power and inefficiency in expressing complex state features, particularly in continuous state-action spaces. With the advent of DRL, HDRL emerged by integrating deep learning (DL) techniques into the HRL framework. This integration not only expands the theoretical foundation of HRL but also leverages deep neural networks to enhance feature extraction and policy learning. Consequently, HDRL exhibits a more efficient and flexible hierarchical structure, making it well-suited for solving complex tasks that were previously intractable using standard HRL methods.

\begin{figure*}[t]
\centering
\includegraphics[width= 0.8\textwidth]{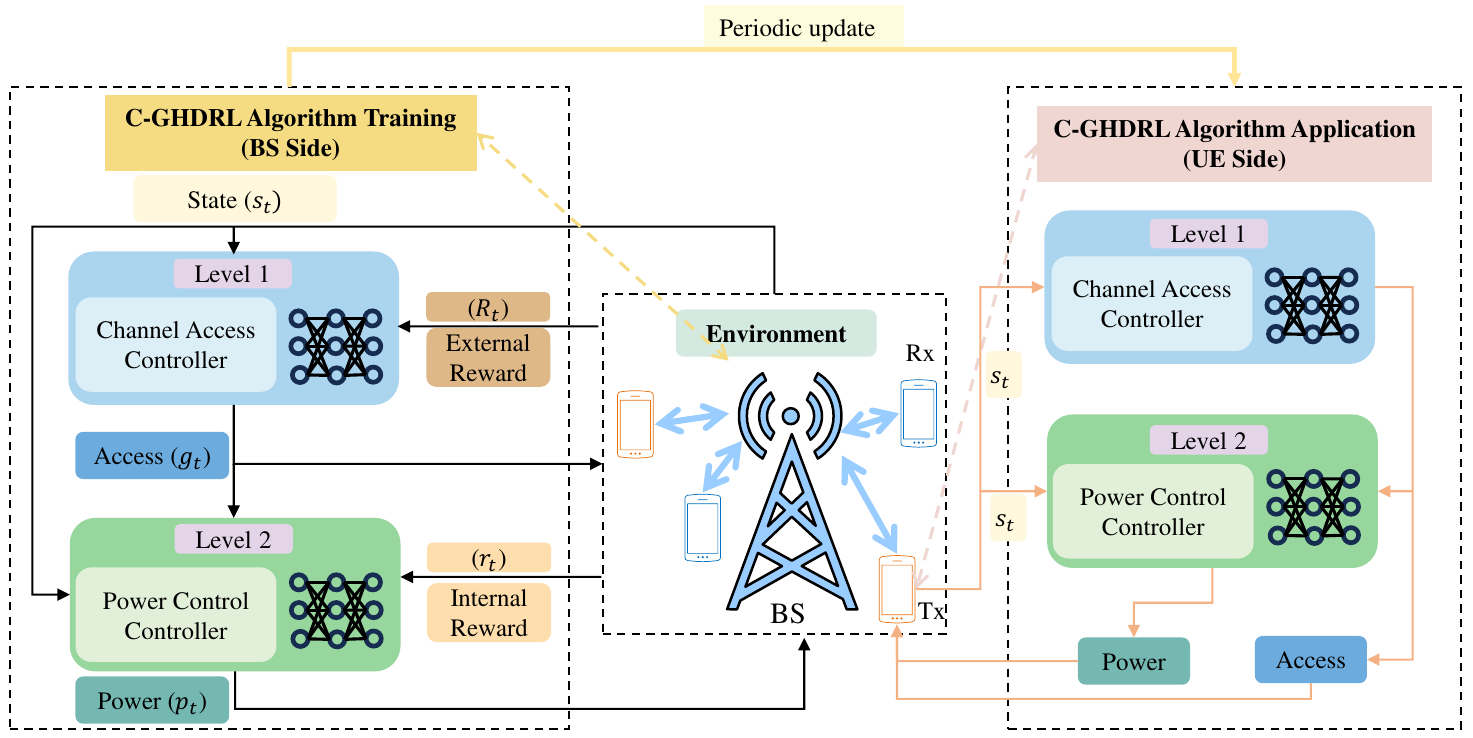}
\caption{Framework of the C-GHDRL algorithm.}
\label{fig:C-GHDRL}
\end{figure*}

Subgoal-based Hierarchical Deep Reinforcement Learning (GHDRL) differs from traditional HRL by not explicitly defining options. Instead, it identifies specific states as sub-goals, represented by the set $\mathcal{G}$. The process of moving from the current state s to a subgoal $g \in \mathcal{G}$ is treated as an option, and an upper-level policy $\pi_h:\mathcal{S} \rightarrow \mathcal{G}$ is learned based on the external reward $\mathcal{R}^{out}$. This approach incrementally guides the agent toward completing the overall task. The design of sub-goals effectively constructs multiple SMDPs, breaking down the original problem into smaller, more manageable tasks and reducing the complexity of the state space. Transitioning between sub-goals signifies the completion of a specific sub-task, providing a crucial mechanism for addressing sequential decision-making processes \cite{ref32}.

Universal Value Function Approximators (UVFA) form the theoretical basis for the GHDRL framework, allowing most classic DRL algorithms to be seamlessly integrated as the underlying algorithms of GHDRL. In this framework, the upper-layer controller takes the current state $s$ as input and outputs a sub-goal $g$. The lower-layer controller then uses the state-subgoal pair $(s,g)$ as an extended state to learn the lower-layer policy $\pi$. The upper-layer policy is trained based on external rewards, while the lower-layer policy is optimized using internal rewards \cite{ref25}. For any extended state $(s,g)$ and internal reward $r^{in}$, the formula for the extended state value function $V_\pi(s,g)$ is defined as:
\begin{equation}
V_\pi(s,g) = \mathbb{E}[\sum\limits_{t=0}^{\infty}r^{in}(s_t,a_t)\prod_{k=0}^t \gamma_g(s_k)|s_0=s],\label{value function}
\end{equation}
where $\gamma_g$ represents the discount coefficient under the subgoal constraint.

Correspondingly, the formula for the extended state-action value function $Q_{pi}(s,g,a)$ is defined as:
\begin{equation}
Q_{pi}(s,g,a) = \mathbb{E}_{s'}[r^{in}(s,a) + \gamma_g(s')V_\pi(s',g)].\label{action value function}
\end{equation}

\begin{description}
\item[B.] Channel Access and Power Control based on C-GHDRL
\end{description}

The joint optimization of channel access and power control can be decomposed into two sub-problems: channel access and power control. This hierarchical structure with multi-step decomposition is particularly well-suited for the GHDRL algorithm. However, in the unlicensed spectrum, SL-U users cannot transmit simultaneously due to competition for channel access, and they are unable to receive timely feedback on the environment, which hinders the GHDRL algorithm model from making globally optimal decisions.

To address this challenge, we developed a C-GHDRL algorithm framework, as illustrated in \Cref{fig:C-GHDRL}. In the framework, the BS collects comprehensive global information through collaborative interactions with SL-U users. The BS utilizes this information to train the C-GHDRL model, which is periodically distributed to all SL-U users to ensure continuous refinement and improvement in their decision-making accuracy. To reduce signaling overhead and latency, each SL-U user independently determines its channel access and transmission power. When an SL-U user has data to transmit, the sensed channel state information is input into the user’s local model to determine channel access and transmission power decisions.

In the framework, the C-GHDRL algorithm comprises two components: a channel access controller and a power control controller. The channel access controller is responsible for optimizing channel access, while the power control controller focuses on power control, utilizing the state information and outputs provided by the channel access controller. Since the state transition probabilities are unknown, we employ a deep Q-learning algorithm to train the policies of both the channel access controller and the power control controller within the C-GHDRL frame \cite{ref33}. Specifically, the power control controller estimates the state-value function $Q_1(s,a;g)$ as follows: 
\begin{equation}
\begin{aligned}
Q_1(s,a;g) &= \max_{\pi_{a g}}\mathbb{E}{[\sum\limits_{t'=t}^{\infty}{\gamma^{t'-t}r_{t'}|s_t = s, a_t = a, g_t = g, \pi_{ag}}]}\\
&= \max_{\pi_{a g}} \mathbb{E}{[r_t + \gamma \max_{a_{t+1}} Q_1(s_{t+1},a_{t+1};g)|s_t=s,}\\
&  {a_t=a,g_t=g,\pi_{ag}]},
\label{Q-value function of controller} 
\end{aligned}
\end{equation}
where $g$ represents the goal of agent in state $s$ and $\pi_{ag}=\mathcal{P}(a|s,g)$ is the action policy. $r_t$ represents the internal reward provided to the power control controller.

\begin{algorithm}[t]
\begin{minipage}{\linewidth}
\begin{center}
\caption{\textbf{:} Learning algorithm base on C-GHDQN}
\begin{algorithmic}[1]
    \STATE Initialize experience replay memories $\{\mathcal{D}_1,\mathcal{D}_2\}$ to capacity $\{\mathcal{C}_1,\mathcal{C}_2\}$ for the power control controller and channel access controller respectively
    \STATE Initialize $\{Q_1,Q_2\}$ network parameters with random weights $\{\theta_1,\theta_2\}$ for the power control controller and channel access controller respectively
    \STATE Initialize target $\{Q'_1,Q'_2\}$ network parameters with random weights $\{ \theta_1^{'},\theta_2^{'}\}$ for the power control controller and channel access controller respectively
    \STATE Initialize exploration probability $\varepsilon_{1,g}=1$ for the power control controller for all goals $g$ and $\varepsilon_2=1$ for the channel access controller.
    \FOR {$i=1,num\_episodes$}
        \STATE Initialize communication system and get start state description $s$
        \STATE Obtain subgoal $g$ through $Q_2$ network according to $\varepsilon$-greedy
        \WHILE {$s$ is \textbf{not} terminal}
            \STATE $R = 0$
            \STATE $s_0 \leftarrow s$
            \WHILE{\textbf{not} ($s$ is terminal \textbf{or} goal $g$ reached)}
                \STATE Obtain $a$ through $Q_1$ network according to $\varepsilon$-greedy
                \STATE Execute $a$ and obtain next state $s'$ and obtain intrinsic reward $r(s,a,s')$ from power control controller
                \STATE Store transition $(\{s,g\},a,r,\{s',g\})$ in $\mathcal{D}_1$
                \STATE Randomly sample mini-batches from $\mathcal{D}_1$ and perform a gradient descent by $L_1(\theta_{1,i})$ with respect to the network parameters $\theta_{1}$
                \STATE Every $C$ steps copy current network weights into target network $\theta'_1 \leftarrow \theta_1$
                \STATE $R \leftarrow R + r$
                \STATE $s \leftarrow s'$
            \ENDWHILE
            \STATE Store transition $(s_0,g,R,s')$ in $\mathcal{D}_2$
            \STATE Randomly sample mini-batches from $\mathcal{D}_2$ and perform a gradient descent by $L_2(\theta_{2,i})$ with respect to the network parameters $\theta_{2}$
            \STATE Every $C$ steps copy current network weights into target network $\theta'_2 \leftarrow \theta_2$
            \IF {$s$ is \textbf{not} terminal}
                \STATE Obtain $g$ through $Q_2$ network according to $\varepsilon$-greedy
            \ENDIF
        \ENDWHILE
    \ENDFOR
\end{algorithmic}
\end{center}
\end{minipage}
\end{algorithm}

Similarly, for the channel access controller, the state-value function $Q_2(s,g)$ is defined as:
\begin{equation}
\begin{split}
& Q_2(s,g) \\
& = \max_{\pi_{g}}\mathbb{E}{[\sum\limits_{t'=t}^{t+N}{R_{t} + \gamma \max _{g'} Q_2(s_{t+N},g')|s_t = s, g_t = g, \pi_{g}}]},\label{Q-value function of meta-controller}
\end{split}
\end{equation}
where $N$ represents the number of time steps between two sub-goals. $g'$ represents the goal of the agent in state $s_{t+N}$, and $\pi_g = \mathcal{P}(g|s)$ represents the policy over goals. $R_t$ is reward signals received from the environment. It is worth noting that the transitions $(s_t, g_t, R_t, s_{t+N})$ generated by $Q_2$ operates at a slower time scale than the transitions $(s_t, a_t, g_t, r_t, s_{t+1})$ generated by $Q_1$.

We represent $Q(s,g) \approx Q(s,g; \theta)$ using a nonlinear function approximator with parameters $\theta$, known as a deep Q-network. Each $Q \in \{Q_1, Q_2\}$ can be trained by minimizing the corresponding loss functions, $L_1(\theta_1)$ and $L_2(\theta_2)$. The loss function for $Q_1$ can be expressed as:
\begin{equation}
L_1(\theta_1,i) = \mathbb{E}_{(s,a,g,r,s')\sim D_1}{[(y_{1,i} - Q_1(s,a;\theta_{1,i},g))^2]},\label{loss functions of controller}
\end{equation}
where $i$ denotes the training iteration number and $y_{1,i}=r+\gamma \max_{a'}Q_1(s',a';\theta_{1,i-1},g)$.

When optimizing the loss function, the parameter $\theta_{1,i-1}$ from the previous iteration remains unchanged. The parameter $\theta_1$ can be optimized using the gradient:
\begin{equation}
\begin{split}
&\nabla_{\theta_{1,i}}L_1(\theta_{1,i}) \\ 
&= \mathbb{E}_{(s,a,r,s'\sim D_1)}{[(r+\gamma \max_{a'}Q_1(s',a';\theta_{1,i-1},g)-}\\
&{Q_1(s,a;\theta_{1,i},g))\nabla_{\theta_{1,i}Q_1(s,a;\theta_{1,i})}]}, \label{gradient of controller}
\end{split}
\end{equation}
The loss function $L_2(\theta_2)$ and its gradients can be derived using a similar procedure.

Consequently, the C-GHDRL algorithm implemented in our framework is built upon the subgoal-based hierarchical deep Q network (GHDQN) model. The fundamental parameters (agent, state, goal, action, reward) in the C-GHDQN algorithm proposed in this paper are as follows: 

1) \emph{Agent}: We propose a centralized optimization method employing BS as an agent. The BS utilizes data from all users to learn about global policy. 

2) \emph{Subgoal}: The subgoal space $\mathcal{G}$ output by the upper layer channel access controller is defined as follows: if the channel is to be accessed, $g(t)=1$; otherwise, $g(t)=0$.

3) \emph{State}: For the upper layer channel access controller, its state is 
\begin{equation}
s_{mc}(t) = [p_s^T(t-1), \boldsymbol{p}_{ch}^R(t)],
\end{equation}
where $p_s^T(t-1)$ represents the power sent by the SL-U user at the last slot, and $\boldsymbol{p}_{ch}^R(t)$ represents the channel energy detected by all SL-U users except itself. For the lower layer power control controller, it is an extended state represented as follows:
\begin{equation}
\begin{split}
s_{c}(t) & = [s_{mc}(t),g(t)]\\
         & = [p_s^T(t-1),\boldsymbol{p}_{ch}^R(t),g(t)],   
\end{split}
\end{equation}

4) \emph{Action}: The action space $\mathcal{A}$ represents the power
adjustment operation. At each slot $t$, the action $a_t \in \mathcal{A}$ can be defined as 
\begin{equation}
    \mathcal{A} = [p_{s,min}^T, \cdots, p_{s,max}^T].
\end{equation}

5) \emph{Reward}: In order to enhance the transmission rate of the coexistence system while maintaining fairness, we define the internal reward $r_t(g)$ and the external reward (environmental reward) $R_t$ as:

\begin{equation}\label{r}
	r_t(g)=\left\{
	\begin{aligned}
		 U & ,  \ \  \eta_{s} \geq \eta_{s,min} \ and \ \eta_{w} \geq \eta_{w,min} \\
		 -U &,  \ \ \ \ \ \ \ \ \ \ \ \ \ \ \ otherwise. 
	\end{aligned}
	\right.
\end{equation}

\begin{equation}\label{R}
R_t=\sum r_t(g).
\end{equation}

We use a deep Q network (DQN) to learn optimal policies for the power control controller and channel access controller. Algorithm 1 shows the specific steps of our C-GHDQN method.

\section{Simulation Result}
In this section, we first present the key parameters of the simulation scenario. Following that, we conduct a performance analysis of the CCHA mechanism. Finally, we evaluate the effectiveness of the C-GHDRL algorithm in addressing the joint optimization of channel access and power control.

\begin{description}
\item[A.] Simulation Settings
\end{description}

We simulate a coexistence scenario in which SL-U and Wi-Fi users operate within a $400 \times 400\ m^2$ environment. In this simulation, a transmission from either SL-U or Wi-Fi users is considered successful when the conditions $\eta_{s,m} \ge \eta_{s,min}$ and $\eta_{w,n} \ge \eta_{w,min}$ are met. The traffic model for users in the coexistence system follows FTP Model 3 \cite{ref33}. For clarity and convenience, the detailed simulation parameters are summarized in Table \ref{table:parameters}.

In our experiments, the deep neural network (DNN) used to approximate the action-value function is composed of three fully connected feedforward hidden layers. The number of neurons in the three hidden layers is 256, 256, and 512, respectively. The Rectified Linear Unit (ReLU) activation function is applied to the first and second hidden layers, and the tanh activation function is used for the third layer. The experience replay memory $\mathcal{D}$ has a capacity $\mathcal{C}$ of 1000, and the Adam optimizer is employed to update the network weights $\theta$, with a minibatch size of 256. The exploration probability $\epsilon$ decreases linearly from 1 to 0 throughout the iterations. The discount factor $\gamma$ is set to 0.8, and the learning rate is $10^{-3}$.

\begin{table}[t]
\center
\caption{Simulation Parameters} 
\scriptsize
\resizebox{0.45\textwidth}{!}{
\begin{tabular}{|c|c|} 
\hline
\makebox[0.2\textwidth][c]{\makecell[c]{Parameters}} & Value \\
\hline
number of SL-U user pairs ($M$) & [20, 24, 28, 32] \\ 
\hline
number of Wi-Fi users ($N$) & [20, 24, 28, 44] \\  
\hline
carrier center frequency ($f_c$) & 5.8 GHz \\  
\hline
channel bandwidth (B) & 20 MHz \\
\hline
sub-carrier space & 30 KHz \\
\hline
channel occupancy time ($T_{cot}$) & 2 ms \\
\hline
power spectral density of AWGN & -174 dBm/Hz \\  
\hline
SIFS & 16 $\mu$s \\  
\hline
DIFS & 34 $\mu$s \\  
\hline
$T_d$ & 34 $\mu$s \\  
\hline
$T_{sl}$ & 9 $\mu$s \\  
\hline
weighted factor$(\beta)$ & 0.8\\
\hline
period of traffic & 10 ms\\
\hline
user's model update time & 100 ms\\
\hline
transport block size & 300 Bytes\\
\hline
$\eta_{s,min}$ & 15 dB \\  
\hline
$\eta_{w,min}$ & 6 dB \\  
\hline 
\end{tabular}
\label{table:parameters}
}
\end{table}

\begin{description}
\item[B.] Performance of the CCHA mechanism
\end{description}

To evaluate the effectiveness of our proposed CCHA mechanism, we compare it with three benchmark schemes. Given that the CCHA mechanism involves two key variables, allocation scheme and channel access. To ensure a comprehensive comparative analysis, we employ the control variable method to design the three benchmark schemes.
\begin{enumerate}[label=\arabic*)] 
\item \textbf{CCHA:} The CCHA mechanism integrates channel access and resource allocation. The specific details are described in Section IV.  

\item \textbf{CCHA-T1:} The only difference between the scheme and the scheme of CCHA is in the channel access mechanism for SL-U users. In the scheme, when SL-U users want to send data, they need to perform Type 1 LBT.

\item \textbf{T12-LBT and DRA:} Unlike the CCHA scheme, in the scheme, resources are independently acquired by each SL-U user, which is referred to as distributed resource allocation (DRA). Additionally, when an SL-U user needs to transmit data, the user first performs Type 1 LBT. Upon successful completion, Type 2 LBT is performed at the designated position before the next SSB.

\item \textbf{T1O-LBT and DRA:} The only difference between this scheme and the scheme of T12-LBT and DRA is that SL-U users do not need to perform a Type 2 LBT before SSB.

\end{enumerate}

\begin{figure}[t]
\centering
\includegraphics[width=0.4\textwidth]{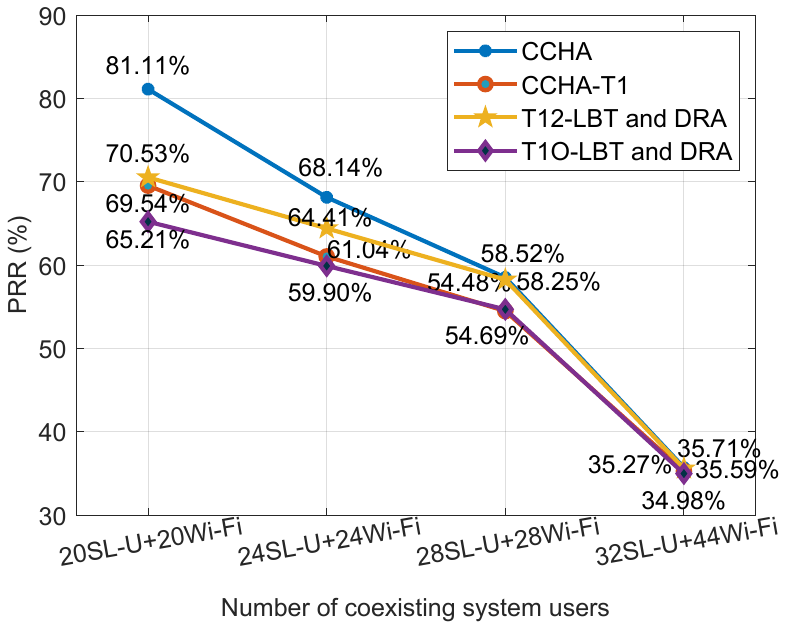}
\caption{Comparisons of PRR over difference schemes.}
\label{fig:traditional PRR}
\end{figure}

\begin{figure}[t]
\centering
\includegraphics[width=0.4\textwidth]{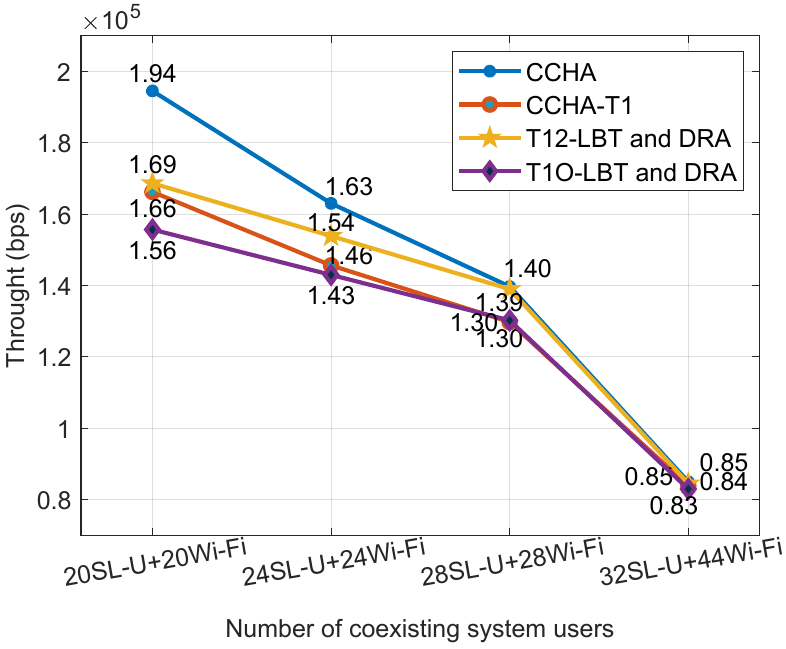}
\caption{Comparisons throught over difference schemes.}
\label{fig:traditional Th}
\end{figure}

In these simulations, we set the transmission power for both SL-U users and Wi-Fi users at 23 dBm. Furthermore, we established a lower channel sensing threshold to reduce the possibility of resource reuse between SL-U users and Wi-Fi users, thereby highlighting the advantages of the CCHA mechanism.

We compare the differences between the scheme proposed in this paper and other schemes in terms of packet reception rate (PRR), throughput and Jain’s fairness index. PRR is the ratio of the number of successfully received packets to the total number of packets sent. Throughput, denoted by $Th$, is defined as the rate of network transmission per unit of time, with the unit expressed in bits per second (bps).
\begin{equation}
Th = \frac{P_{success} \times TB_{size}}{Time}, \label{Th}
\end{equation}
where $P_{success}$ is the number of successfully received packets by SL-U.  The $TB_{size}$ is the transport block size. $Time$ is simulation time.

\begin{figure}[t]
\centering
\includegraphics[width=0.4\textwidth]{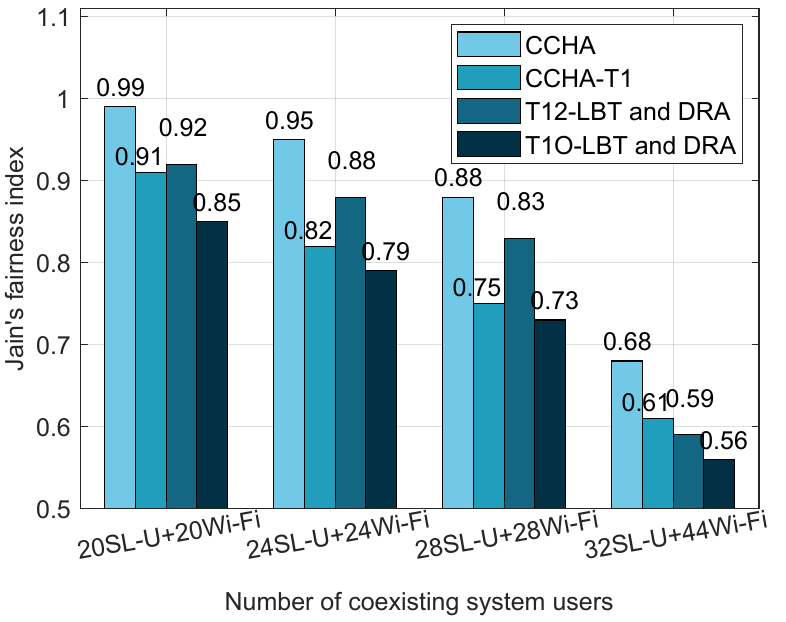}
\caption{Comparisons of Jain's fairness index over difference schemes.}
\label{fig:traditional fairness}
\end{figure}

\Cref{fig:traditional PRR} illustrates the PRR results for different numbers of users in the system. As shown in the figure, the PRR for all schemes decreases as the number of users increases. The decrease is caused by system resources becoming increasingly scarce as the number of users increases. Furthermore, if the number of users increases to a certain scale, the performance deteriorates sharply due to severe resource contention. Under moderate user density, the CCHA scheme outperforms the other schemes in terms of PRR. The reasons include two aspects. On the one hand, the advantages of the two types of channel access are fully utilized in the CCHA mechanism. Type 1 LBT performed by the base station can sense whether the channel is idle initially, while Type 2 LBT performed by the SL-U user can sense whether the channel is idle again before SSB, with a shorter sensing time compared to Type 1 LBT. Therefore, the mechanism effectively improves the successful acceptance rate of data packets. On the other hand, CCHA uses centralized resource allocation (CRA), which can ensure that the resources in the SL-U system are completely orthogonal. This method effectively prevents collisions arising from multiple SL-U users selecting the same resources for data transmission due to the absence of a coordination mechanism in the DRA, further increasing the likelihood of conflicts and reducing the successful packet acceptance rate. Additionally, the PRR performance of the T1O-LBT and DRA schemes is the worst. The T1O-LBT scheme requires SL-U users to perform only Type 1 LBT without reconfirming whether the channel is idle before access. This limitation greatly increases the possibility of collisions with users who have already occupied the channel, leading to a higher rate of packet transmission failure. Furthermore, the absence of a coordination mechanism in the DRA scheme leads to increased potential for collisions among SL-U users, further diminishing the number of successfully transmitted packets.

\Cref{fig:traditional Th} illustrate the throughput results for varying numbers of users in the system. The throughput performance trend in the figure mirrors is similar to the PRR. This is because the CCHA solution enhances resource utilization and mitigates collision issues, resulting in better throughput performance in the coexistence system compared to the three benchmark solutions.

\Cref{fig:traditional fairness} illustrates the comparison of Jain’s fairness index across the four schemes as the number of system users increases. As shown in the figure, Jain’s fairness index decreases for all schemes as the number of users rises. The decrease is caused by system resources becoming increasingly scarce as the number of users rises. The CCHA scheme outperforms the other comparison schemes regarding the Jain fairness index. This is because the CCHA mechanism leverages the advantages of two types of channel access and centralized resource allocation, effectively ensuring fairness while enhancing resource utilization. In addition, the T1O-LBT and DRA schemes exhibit the worst Jain fairness index performance. This is because SL-U users only perform Type 1 LBT and do not reconfirm the channel status before accessing the channel. This limitation increases the likelihood of collisions with Wi-Fi users, leading to a lower fairness index.

\begin{description}
\item[C.] Performance of the Proposed C-GHDRL
\end{description}

As the number of users increases, system performance degrades significantly due to resource constraints. To address this, SL-U and Wi-Fi users are allowed to perform appropriate resource reuse, improving the overall performance of the coexistence system. To achieve this, we employ the C-GHDRL algorithm to jointly optimize the channel access and transmission power for SL-U users, further enhancing system performance. In this section, the numbers of SL-U and Wi-Fi users are set to 32 and 44, respectively. Additionally, the transmission power of SL-U users is set between $-40$ dBm and 0 dBm due to the relatively short distance between their transmitters and receivers. Meanwhile, Wi-Fi users dynamically adjust their transmission power using a closed-loop power control strategy to adapt to the dynamic variations in the channel environment \cite{addref45}. 

To verify the performance of the C-GHDRL algorithm, we utilize four schemes for comparison, i.e., the DQN algorithm, the fundamental scheme of CCHA (F-CCHA), random algorithm, and open-loop power control algorithm.

\begin{enumerate}[label=\arabic*)] 
\item \textbf{C-GHDRL algorithm:} The C-GHDRL algorithm-based scheme builds upon the F-CCHA scheme and leverages the C-GHDRL algorithm to dynamically adjust channel access and transmission power for SL-U users. For further details, refer to Section V.

\item \textbf{DQN algorithm \cite{addref46}:} The DQN algorithm-based scheme builds upon the F-CCHA scheme and leverages the DQN algorithm to dynamically adjust transmission power for SL-U users.

\item \textbf{F-CCHA:} The F-CCHA scheme acts as a benchmark scheme that employs the CCHA mechanism for SL-U users. In the scheme, the LBT energy sensing threshold for SL-U users is fixed, and their transmission power is 0 dBm.

\item \textbf{Random algorithm \cite{addref47}:} The random algorithm-based scheme builds upon the F-CCHA scheme and leverages the random algorithm to dynamically adjust channel access and transmission power for SL-U users.

\item \textbf{Open-loop power control algorithm \cite{addref48}:} The open-loop power control algorithm-based scheme builds upon the F-CCHA scheme and leverages the open-loop power control algorithm to dynamically adjust transmission power for SL-U users.

\end{enumerate}

\begin{figure}[t]
\centering
\includegraphics[width=0.4\textwidth]{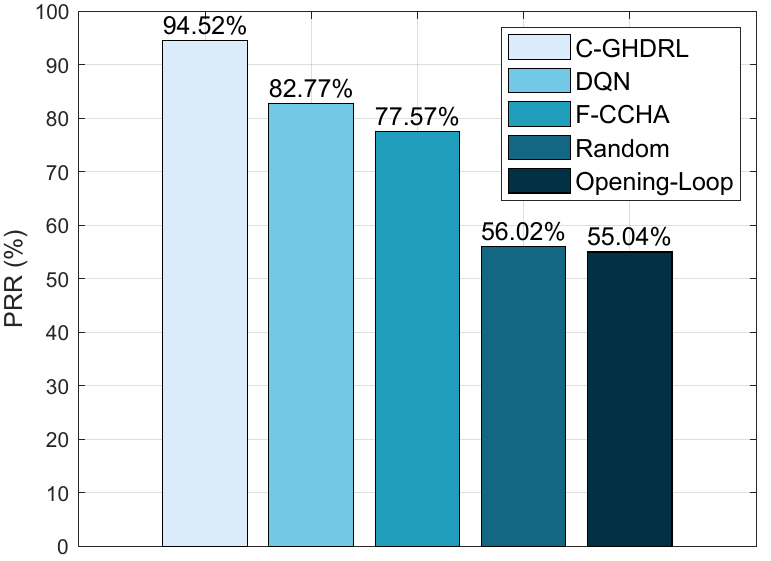}
\caption{Comparisons the PRR over difference schemes.}
\label{fig:PRR-AI}
\end{figure}
\begin{figure}[t]
\centering
\includegraphics[width=0.4\textwidth]{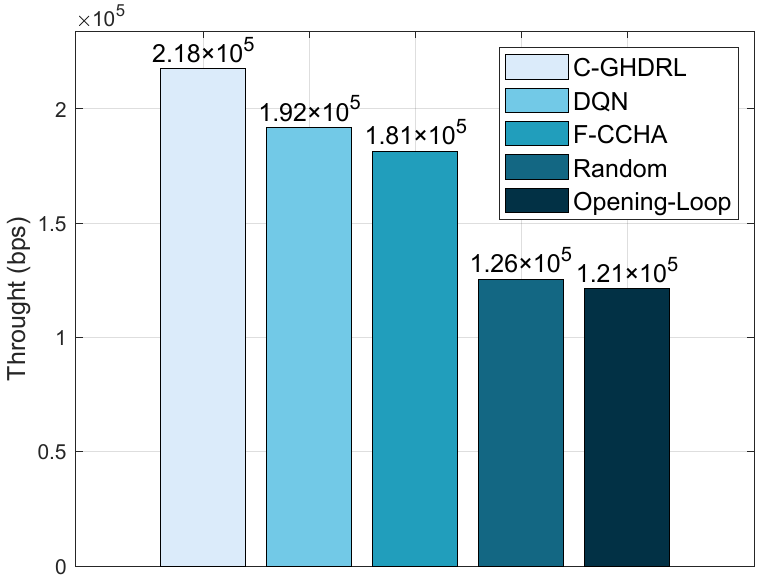}
\caption{Comparisons the throught over difference schemes.}
\label{fig:Throught -AI}
\end{figure}

\Cref{fig:PRR-AI} shows the PRR performance of five different schemes. As depicted in the figure, the PRR of the scheme utilizing the C-GHDRL and the DQN algorithm is higher than the benchmark scheme of F-CCHA. Notably, the scheme employing the C-GHDRL algorithm exhibits the highest PRR. This is because adjusting the transmission power and channel access opportunities of SL-U users can effectively increase the transmission success rate of both SL-U users and Wi-Fi users. In addition, since the scheme based on the DQN algorithm only adjusts the transmission power of SL-U users, its PRR is lower than the scheme based on the C-GHDRL algorithm. From the figure, we can also observe that both the random scheme and the scheme based on open-loop power control have lower PRR than the benchmark scheme of F-CCHA. This is because the two schemes have shortages in adjusting the transmission power and channel access opportunities of SL-U users. The random scheme involves randomness in both transmission power and access channel. Although the scheme of open-loop power control can determine the transmission power by calculating the path loss based on the reference signal, it cannot estimate the interference power in the channel. Therefore, it is difficult for the power adjustment strategy in the open-loop power control scheme to calculate a transmission power that guarantees successful data transmission for SL-U users.

\Cref{fig:Throught -AI} shows the throughput performance of various schemes. The reasons and phenomena are the same as PRR in \Cref{fig:PRR-AI}.

\begin{figure}[t]
\centering
\includegraphics[width=0.4\textwidth]{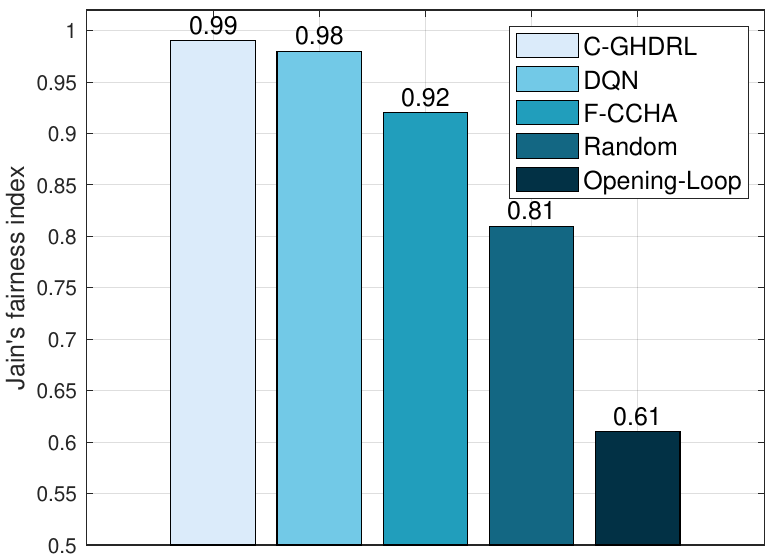}
\caption{Comparisons the Jain's fairness index over difference schemes.}
\label{fig:fairness-AI}
\end{figure}

\Cref{fig:fairness-AI} shows the comparison of Jain’s fairness index of five different schemes. As can be seen from the figure, Jain’s fairness index of the schemes based on the C-GHDRL and the DQN algorithm is higher than the benchmark scheme of F-CCHA. The phenomenon shows that schemes based on the C-GHDRL and DQN algorithms can improve the performance of coexistence systems. However, the scheme based on the DQN algorithm improves system performance less than the scheme based on the C-GHDRL algorithm. This is because the scheme based on the C-GHDRL algorithm effectively ensures the coexistence of SL-U users and Wi-Fi users by adjusting the transmission power and channel access opportunities of SL-U users, thereby achieving the highest Jain’s fairness index. The scheme based on the DQN algorithm cannot control channel access opportunities, which has a certain impact on Wi-Fi users, resulting in a slight decrease in Jain’s fairness index. Furthermore, Jain’s fairness index of the random scheme and the open-loop power control-based scheme is lower than the benchmark scheme of F-CCHA. This is because the random scheme is unpredictable regarding transmission power and channel access opportunities. Therefore, the impact on Wi-Fi users cannot be controlled, resulting in a low Jain’s fairness index. Based on the open-loop power control scheme, SL-U users cannot detect interference in the channel, which seriously affects the performance of the coexistence system and makes Jain’s fairness index of this scheme the smallest.

\section{Conclusion}
In this paper, we investigate methods to enhance the overall performance of coexistence systems operating in the unlicensed spectrum while ensuring fair coexistence between SL-U users and Wi-Fi users. First, we propose a CCHA mechanism that integrates channel access with resource allocation. This mechanism enhances spectrum fairness in coexistence systems while improving spectrum utilization. Next, we develop a Cooperative Sub-goal-based Hierarchical Deep Reinforcement Learning (C-GHDRL) framework to further optimize spectrum utilization. The framework enables joint optimization of channel access and power control through cooperative interaction between the BS and SL-U users, thereby significantly improving the overall performance of the coexistence system. Finally, we formulate a joint mathematical model for channel access and power control, aiming to balance fairness and transmission rate by defining an appropriate reward function within C-GHDRL algorithm. Future research will focus on exploring more efficient collaborative mechanisms under conditions of incomplete global information.

\bibliographystyle{IEEEtran}
\renewcommand{\bibfont}{\footnotesize}
\bibliography{ref.bib} 
\end{document}